\documentclass[twocolumn]{fairmeta}

\usepackage{amsmath}
\usepackage{amssymb}
\usepackage{enumitem}
\usepackage{balance}

\newcommand\R{\mathbf{R}}

\title{CMSL: Constructive Multi-Sequence Learning for Recommendation Systems}

\author[1,*]{Zikun Cui}
\author[1,*]{Renzhi Wu}
\author[1,*]{Junjie Yang}
\author[1,*]{Li Sheng}
\author[1,*]{Jijie Wei}
\author[1,*]{Linfeng Liu}
\author[1,*]{Tai Guo}
\author[1,*]{Tao Jia}
\author[2,*]{Xiaodong Wang}
\author[1,*]{Hong Li}
\author[3,*]{Li Yu}
\author[1,*]{Sri Reddy}
\author[1,*]{Hong Yan}

\affiliation[1]{Meta MRS}
\affiliation[2]{Meta PyTorch}
\affiliation[3]{Meta FB Monetization}

\contribution[*]{All authors contributed equally to this research}

\abstract{Sequence learning has emerged as the promising paradigm in recommendation systems, surpassing traditional Deep Learning Recommendation Models (DLRM) by capturing the temporal nuances of user behavior. However, current state-of-the-art architectures operate under a limiting analogy: they treat user history as a monolithic chronological sequence like a sentence in a Large Language Model (LLM).
We observe a fundamental divergence between natural language and recommendation data: unlike the linear, logical flow of text, user history is inherently multi-faceted. A user's journey is a fragmented reflection of diverse interests, resulting in much weaker coherence between items than is found in LLM training data. This lack of structural unity leads to context pollution. In single-sequence modeling, unrelated behaviors compete for the same attention budget. This ``noisy'' signal dilutes the model's focus, effectively capping its ability to discern high-intent patterns from background activity.
To address this, we propose Constructive Multi-Sequence Learning (CMSL), a paradigm shift from passive sequence ingestion to active ``context engineering'' that constructs multiple coherent sequences in latent space.
CMSL leverages a learnable Sequence Construction Module to disentangle user history into ``pure'' thematic strands, followed by a linear attention mechanism to efficiently model these strands at scale.
CMSL has been deployed across ranking and retrieval tasks and across four major surfaces at Meta.}

\date{\today}
\correspondence{\email{renzhiwu@meta.com}}

\metadata[Keywords]{Recommendation systems, Sequence learning, Generative sequence modeling}

\begin{document}

\maketitle

\section{Introduction}
Building on the success of Transformers in language and vision tasks, sequence learning has emerged as a powerful paradigm for recommendation systems. Unlike the previous DLRM approaches~\citep{covington2016deep,cheng2016wide,zhou2018deep,tang2020progressive,wang2021dcn,mudigere2022software,xia2023transact,zhang2024wukong}, which rely heavily on aggregated categorical features and often loses sequential information, sequence learning retains important insights into the user's journey.
A large body of work has developed increasingly powerful sequential recommenders, demonstrating notable success~\citep{kang2018self, zhai2024actions, zhang2025onetrans, zeng2025interformer, deng2025onerec}.

A notable trend is that advances in sequential recommendation have increasingly been inspired by LLM research~\citep{kang2018self, zhai2024actions, zhang2025onetrans, zeng2025interformer, deng2025onerec}. Many RecSys models adopt a similar interface: represent a user's interaction history as a sequence of tokens (item IDs), feed it into a self-attention backbone, and treat next-item prediction as an autoregressive continuation problem. This ``history-as-prompt'' view is attractive because it provides a unified modeling recipe and a straightforward path to scaling with model size and context length.

\noindent\textbf{The Context Pollution Problem:}
While a sentence is governed by strict syntactic rules and linear semantic dependencies, a user's digital footprint is a ``noisy mosaic.'' It is a disjointed mixture of overlapping interests, exploratory clicks, and accidental interactions. Because recommendation data lacks the inherent narrative unity found in text, direct applications of LLM-style sequence modeling often struggle to capture the underlying user intent effectively.

\begin{figure*}[t]
  \centering
  \captionsetup{justification=centering,singlelinecheck=true}
  \includegraphics[width=0.6\textwidth]{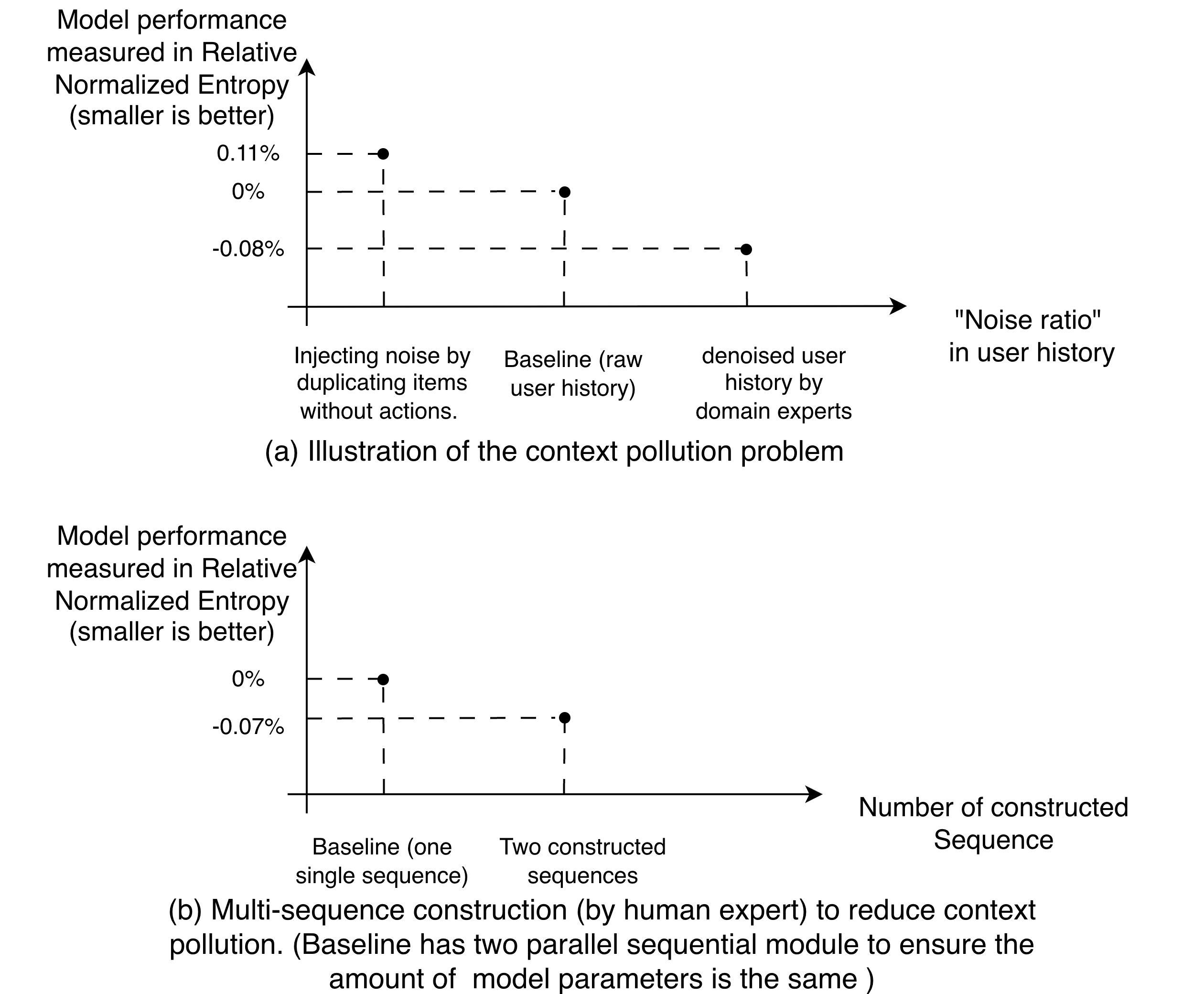}
    \caption{The context pollution problem and constructing multiple coherent sequences to mitigate the issue.}
  \label{fig:context_pollution}
\end{figure*}


Irrelevant or weakly related events compete for attention and distort the representation needed for the current prediction. Context Pollution is a well-known failure mode in LLMs, where irrelevant tokens in a prompt degrade reasoning and induce hallucination~\citep{shojaee2025illusion, mei2025survey, hong2025context}. In recommendation, the problem can be even more acute because pollution is not only random noise but also semantic interference. For example, forcing the model to compute dependencies between a user's professional learning (e.g., ``coding tutorials'') and casual entertainment (e.g., ``cooking videos'') encourages spurious correlations and creates a representation bottleneck. Attention is drawn toward irrelevant events, diluting the signal of the user's true current intent and leading to suboptimal predictions.
Figure~\ref{fig:context_pollution}(a) empirically illustrates this phenomenon: injecting low-signal events into a user's history degrades performance, while expert-driven denoising improves it. Despite its prominence in LLM discussions, context pollution remains under-explored in sequential recommendation.


\noindent\textbf{From ``modeling a sequence'' to ``constructing context''.}
We pose a critical question: \textit{Is there a better formulation of sequence learning for recommendation systems --- one that treats user history not as a ``raw prompt'' to be read, but as a context to be constructed?}

We propose \textit{Constructive Multi-Sequence Learning} (CMSL), a new paradigm deployed in our massive-scale production systems across major organic and ads surfaces.
Instead of accepting user history as an arbitrary, monolithic sequence, CMSL actively constructs multiple coherent latent sequences from the set of user interactions.
Each latent sequence performs self-attention separately, reducing cross-intent interference and mitigating context pollution.
As illustrated in Figure~\ref{fig:context_pollution}(b), even an expert-constructed split into two coherent sequences can outperform a single-sequence baseline.

This approach aligns with the philosophy of ``context engineering'' in LLMs: the idea that performance depends on the curation and organization of the context provided to the model. However, recommendation requires a domain-specific adaptation: unlike words, item IDs are abstract tokens without explicit semantics, making manual curation infeasible.
Therefore, CMSL performs Implicit Context Engineering through a learnable module that operates in the latent space to automatically disentangle the noisy raw user history into multiple latent sequences.
These sequences act like ``views'' of user history, emphasizing different aspects of relevance and reducing semantic interference.
Our key contributions are:
\begin{itemize}[leftmargin=*]
    \item \textbf{A New Perspective on Sequential Recommendation}: We highlight the ``Context Pollution'' problem in sequential recommendation, and propose a multi-sequence construction and learning paradigm to tackle the problem.
    \item \textbf{Multi-sequence Construction Module}: We propose a learnable module that transforms raw history into multiple latent streams.
    This module leverages cross attention to perform intent-aware construction allowing the model to adaptively curate context at query time.
    \item \textbf{Scalable Linear Attention for Multi-sequence Modeling}: We develop an efficient linear-time attention mechanism that reduces the computational cost of Transformer-style attention, enabling efficient modeling of constructed multi-sequences.
    \item \textbf{Industrial-Scale Validation}: We implemented CMSL and applied it to industry-level data with billions of daily active users, demonstrating the effectiveness of our approach.
\end{itemize}

\section{Related Works}
Recent industrial sequential recommendation backbones emphasize context-conditioned sequence modeling and scalability: HSTU~\citep{zhai2024actions} improves relevance modeling via target-aware attention so the encoder can focus on history most pertinent to the current context, while methods such as OneTrans~\citep{zhang2025onetrans} and Longer~\citep{chai2025longer} explore architectural and efficiency improvements to make long-history modeling practical at scale.
In parallel, InterFormer~\citep{zeng2025interformer} and related heterogeneous interaction models focus on better combining sequential signals with non-sequential features (user/item/surface context), arguing that richer cross-modal interaction is critical for expressive representations. Finally, OneRec~\citep{deng2025onerec, zhou2025onerec} reflects a growing trend of leveraging powerful generative sequence modeling paradigms to unify traditional cascaded system into one end-to-end model.
Most of these methods still share a common formulation: represent the user's past as one monolithic chronological sequence, and rely on stronger attention/fusion or longer context windows to extract what matters. Our work is complementary: we target a different bottleneck, i.e. context pollution in long, heterogeneous histories, by constructing multiple context-conditioned latent sequences from the same raw history before applying sequence attention. This enables the model to separate competing intents into distinct streams, and improve model learning.

\begin{figure*}[t]
  \centering
  \captionsetup{justification=centering,singlelinecheck=true}
  \includegraphics[width=0.80\textwidth]{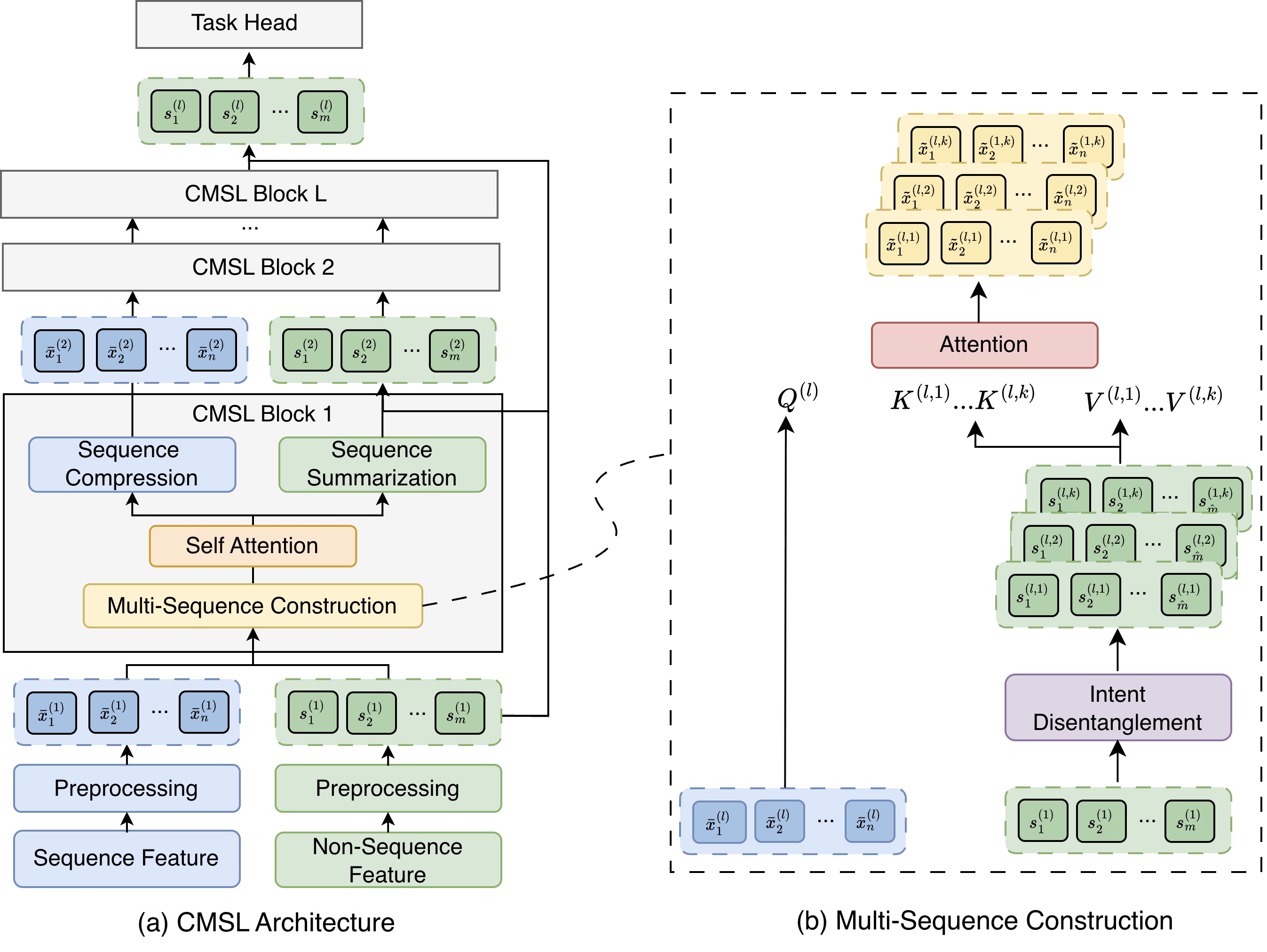}
    \caption{Overall Model Architecture.}
  \label{fig:cmsl_figure}
  \vspace{-3mm}
\end{figure*}

\section{Methodology}
This section details the proposed Constructive Multi-Sequence Learning (CMSL) framework. The overall architecture is shown in Figure~\ref{fig:cmsl_figure}. We first introduce the feature preprocessing in Section 3.1, followed by the CMSL model architecture in Section 3.2.

\subsection{Feature Preprocessing}

\subsubsection{Non-Sequence Features}
Non-Sequence features can be categorized into dense feature and sparse features. We apply different preprocessing to transform these raw features into d-dimension embedding that can be used in the model.
\begin{itemize}[leftmargin=*]
    \item \textbf{Dense Features:} Continuous number such as user age, product price, and video watch time are examples of dense features. These raw numbers are concatenated into a  raw feature vector $\mathbf{v}_{dense} \in \R^m$. To ensure numerical stability and consistent scaling across different feature distributions, this vector is normalized and then projected into a $d$-dimensional embedding space:
    $\mathbf{S}_{dense} = \mathbf{W}_d \cdot \text{Norm}(\mathbf{v}_{dense})$
    where $\mathbf{W}_d \in \R^{d \times m}$ is a learnable weight matrix.
    \item \textbf{Sparse Features:} Example of sparse features include country code, user language, product category, etc. These features typically has a high cardinality are processed using an embedding-based approach. Each sparse feature $j$ is associated with a unique embedding table $\mathcal{T}_j \in \R^{V_j \times d}$. For features containing multiple values (e.g., a list of interest tags), we use the summation of the embeddings as the final embedding. Each sparse feature will be transformed into a $d$-dimensional embedding, which is the same as dense feature.
\end{itemize}

\subsubsection{Sequence Features}
We define the interaction history of a user $u$ as a set of sequences $\mathcal{X} = \{X^{(1)}, X^{(2)}, \dots, X^{(B)}\}$. Each sequence $X^{(b)} = \{x^{(b)}_1, x^{(b)}_2, \dots, x^{(b)}_n\}$ represents a specific interaction modality (e.g., clicks, likes, and comments). Each token $x^{(b)}_i$ corresponds to an item or action. Each sequence $X^{(b)}$ is processed at the token level through a shared embedding layer, which transform a sequence of token into a sequence of embedding.

In large-scale social media platform, each user typically possesses a vast number of interaction sequences. Processing all these sequences independently in the model introduces a prohibitive computational overhead. To mitigate this computational cost while preserving model performance, we group the sequences by shared attributes and then compress them into a single sequence (e.g. click, like, and comments sequences on the same product surface). For a group of related sequences $\mathcal{X}_{g} = \{X^{(1)}, \dots, X^{(M)}\}$, each sequence first passes through a sequence-specific MLP to calculate a score $s^{(m)}$ for each token:
$s^{(m)}_i = \sigma(\text{MLP}^{(m)}(x^{(m)}_i))$
where $\sigma$ is the sigmoid activation function. The original embeddings are then weighted by these scores and summed across the sequence dimension to form a single compressed sequence:
$\bar{X} = \sum_{m=1}^{M} (s^{(m)} \cdot X^{(m)}) \in \R^{n \times d}$.
By consolidating related sequence through this gated summation, we significantly reduces computation cost in subsequent layers without discarding the underlying interaction information.


\subsection{CMSL Block}
The core architecture of CMSL consist of $N$ layer of stacked CMSL blocks (Figure~\ref{fig:cmsl_figure}).
Each block refines the sequence representation while explicitly controlling context interference. Concretely, each block contains four components:
\begin{itemize}[leftmargin=*]
    \item Multi-sequence construction: construct $K$ context-conditioned sequences from the same input history.
    \item Self-attention: model within-sequence dependencies separately inside each constructed sequence.
    \item Sequence summarization: extract compact vectors summarizing each constructed sequence for cross-layer/context propagation.
    \item Sequence compression: merge the $K$ sequences back into one sequence to avoid sequence-count explosion across layers.
\end{itemize}
This design preserves the benefits of sequence modeling, while preventing unrelated intents from competing inside a single attention context.

\subsubsection{Multi-Sequence Construction}
The multi-sequence construction module is the key departure from monolithic sequential modeling (Figure~\ref{fig:cmsl_figure}(b)).
Intuitively, the same raw history can be ``read'' through different contextual lenses. CMSL learns these lenses and uses them to construct $K$ contextual coherent latent sequences.

\noindent\textbf{Step 1: intent disentanglement on non-sequence features.}
At layer $l$, we first project the non-sequence representations into $K$ context-specific sets $S^{(l,k)}$ with $K$ projection heads. This disentangles user intents into $K$ different contexts.
Each set $S^{(l,k)}$ is then projected into a key $K^{l,k}$ and a value $V^{l,k}$.

\noindent\textbf{Step 2: sequence construction with Attention.}
For each pair of key $K^{l,k}$ and value $V^{l,k}$, we project the current sequence representation $\bar{X}$ to form the query $Q^{l}$. After that, multi-head attention is applied to generate a new sequence $\tilde{X}^{(l,k)}$.


By these two steps, we construct $K$ sequences focusing on different aspects of context.

\subsubsection{CMSL Attention}
After multi-sequence construction, each sequence $\tilde{X}^{(l,k)}$ undergoes a self attention transformation to capture internal temporal correlation within that specific context, instead of forcing a single attention module to reconcile conflicting intents.
Sequence self attention is a heavy operation because it's complexity quadratically increase with sequence length.
Multi-sequence construction further increase the computation cost by $K$ times.
To reduce computation cost, we use a linear-time attention mechanism that approximate the HSTU~\citep{zhai2024actions} attention that widely used in Meta's recommendation system.

Unlike vanilla transformers, HSTU attention mechanism does not have row-wise normalization and computes the output as
\begin{equation}
    \mathrm{SiLU}\left(M \circ Q K^\top\right) V
\end{equation}
where $\mathrm{SiLU}(\cdot)$ is the sigmoid linear unit function and $M \in \{0, 1\}^{n \times n}$ be a attention mask which could be the causal mask. Our linear-time attention mechanism is motivated by the following observations:

\begin{equation}
\label{eq:silu}
    \mathrm{SiLU}(x^Ty) \approx \frac{1}{2}(x^Ty) + \frac{1}{4}(x^Ty)^2 - \frac{1}{48}(x^Ty)^4 +  \ ...
\end{equation}

\begin{equation}
\resizebox{\columnwidth}{!}{$\phi(x) = [x_1, x_2, \dots, x_d, x_1x_2, \dots, x_i x_j, \dots, x_d x_d] \in \R^{d^2+d}$}
\end{equation}
Equation~\ref{eq:silu} is the Taylor series of SiLU activation. $\phi(x)$ is a degree-2 polynomial kernel of $x$.
It can be shown that:
\begin{equation}
\label{eq:silu_app}
    \phi(x)^\top \phi(y) \le \mathrm{SiLU}\left(x^\top y\right) \le \phi(x)^\top \phi(y) + O\left(\left(x^\top y\right)^3\right)
\end{equation}

Equation~\ref{eq:silu_app} implies that we can approximate the HSTU attention, $\mathrm{SiLU}(Q K^\top) V$, by a linear attention $\phi{(Q)}\phi{(K)}^TV$ plus $A V$, where A is a sparse matrix that only contains large entries in $\mathrm{SiLU}(Q K^\top)$. That is, we have:
\begin{displaymath}
\mathrm{SiLU}(Q K^\top) V \approx \phi(Q) \phi(K)^\top V + A V
\end{displaymath}
Therefore, we use linear attention for $\phi(Q) \phi(K)^\top V$. For the large entries ($AV$), inspired by the work of native sparse attention~\citep{yuan2025native}, we developed a kernel to  captures large entries in $QK^T$ matrix.

\subsubsection{Sequence Summarization}
After self attention, we employ Pooling by Multi-head Attention (PMA)~\citep{lee2019set} to summarize sequences into a fixed-length vector.
Sequence summarization is a vital stage in the CMSL architecture, serving to distill highly informative global signals from the fine-grained temporal representations. Without an explicit summarization module, information transfer between layers would be susceptible to the noise inherent in raw interaction sequences. By utilizing the current contextual non sequence feature $S^{(l,k)}$ as a query, PMA selectively extracts behavior patterns that are most relevant to the user's current state, effectively suppressing redundant or irrelevant historical data.


\begin{equation}
    h^{(l,k)} = \text{PMA}(Q=S^{(l,k)}, K=\bar{X}^{(l,k)}, V=\bar{X}^{(l,k)})
\end{equation}
The summarized representations $h^{(l,k)}$ across all sequences will be a part of non-sequence features in the next layer.

\subsubsection{Sequence Compression}
To prevent the number of sequences from growing by a factor of $K$ at every layer $l$,
we concatenate the $K$ sequences and project back into a single latent sequence:
\begin{equation}
    \resizebox{\columnwidth}{!}{$\bar{X}^{(l)} = \text{MLP}([\bar{X}^{(l-1,1)} \parallel \bar{X}^{(l-1,2)} \parallel \dots \parallel \bar{X}^{(l-1,K)}]) \in \R^{n \times d}$}
    \label{eq:seq_compression}
\end{equation}
This allows the model to learn deep, hierarchical interactions across $L$ layers while maintaining a constant sequence count per layer.

\subsubsection{Final Fusion}
The final representation $H$ is a concatenation of the sequence summarization $H^{b} = [h^{(l,1)} \parallel \dots \parallel h^{(l,k)}]$ from the final layer. This representation is passed to multiple task-specific heads to perform CTR/CVR predictions.

\section{Experimental Results}

We evaluate CMSL on both Ranking and Retrieval tasks.
\begin{itemize}[leftmargin=*]
    \item For ranking, we evaluate with the Normalized Entropy (NE)~\citep{he2014practical} metric of the produced predictions. Smaller NE means better prediction performance. The NE metric is a widely accepted metric at Meta and for ranking models NE is known to be very consistent with A/B test metrics.
    \item For retrieval, since retrieval models often has strong offline-online inconsistency, we directly evaluate the performance with online A/B test.
\end{itemize}

\noindent\textbf{Training Infrastructure}
We train the models on an internal industrial training platform with mixed precision on 128 NVIDIA H100 GPUs and with customized kernel optimizations for memory and throughput efficiency.
Each experiment is trained on 30-50 billion examples sampled from the production traffic.

\subsection{Ranking Evaluation}
We consider two methods to evaluate the effectiveness of CMSL for CTR prediction on four large scale surfaces:
\begin{itemize}[leftmargin=*]
    \item On surface 1, we directly train a CMSL model for engagement prediction and compare with production baseline.
    \item On surface 2, surface 3 and surface 4, we introduce a two-stage architecture designed to decouple heavy computation from the primary ranking path. By offloading the long-context CMSL models to an asynchronous upstream user model, we significantly reduce latency while maintaining high-fidelity signal extraction..
\end{itemize}

We report the NE change (\%) of trained CMSL model over production baseline in Table~\ref{tab:cmsl_ne} on two most important engagement prediction tasks. The table shows that CMSL delivers strong NE gain (0.3\%-0.6\%) in both training and evaluation. Note that >0.03\% NE improvement is considered statistically significant improvement.
\begin{table}[ht]
  \centering
  \caption{CMSL NE change (\%) over production baseline on surface 1.}
  \label{tab:cmsl_ne}
  \begin{tabular}{@{}lll@{}}
    \toprule
    & Comment  & Like \\
    \midrule
    Train NE & -0.54 & -0.31\\
    Eval NE & -0.62 & -0.33\\
  \bottomrule
\end{tabular}
\end{table}

We evaluate the efficacy of the two-stage CMSL architecture by measuring its direct impact on NE within the online production model in Table~\ref{tab:cmsl_ne_feature} on two most important tasks of surface 2/3/4.
\begin{table}[ht]
\caption{NE change (\%) after adding CMSL user embeddings into over production model on three surfaces.}
\label{tab:cmsl_ne_feature}
\centering
\resizebox{\columnwidth}{!}{%
\begin{tabular}{|l|cc|cc|cc|}
\hline
 & \multicolumn{2}{c|}{Surface 2} & \multicolumn{2}{c|}{Surface 3} & \multicolumn{2}{c|}{Surface 4} \\ \cline{2-7}
 & \multicolumn{1}{c|}{CTR} & CVR & \multicolumn{1}{c|}{CTR} & CVR & \multicolumn{1}{c|}{CTR} & CVR \\ \hline
Eval NE & \multicolumn{1}{c|}{-0.12} & -0.10 & \multicolumn{1}{c|}{-0.09} & -0.06 & \multicolumn{1}{c|}{-0.10} & {-0.13} \\ \hline
\end{tabular}%
}
\end{table}

\subsection{Retrieval Evaluation}
To evaluate the effectiveness of CMSL in retrieval, we perform online A/B test on surface 5. We report 4 important engagement metrics in Table~\ref{tab:cmsl_retrieval}, which shows CMSL has statistical significant improvements on all of the four metrics.

\begin{table}[ht]
  \centering
  \caption{A/B test metrics of U2U retrieval with CMSL user embeddings on surface 5.}
  \label{tab:cmsl_retrieval}
  \begin{tabular}{@{}lllll@{}}
    \toprule
    & Metric 1  & Metric 2 & Metric 3 & Metric 4 \\
    \midrule
     & +0.116\% & +0.158\% & +0.171\%  & +0.092\%\\
  \bottomrule
\end{tabular}
\end{table}

\section{Conclusion}
We proposed Constructive Multi-Sequence Learning (CMSL), a framework that improves sequential recommendation by replacing monolithic history modeling with learned context construction. CMSL constructs multiple context-conditioned latent sequences from the same raw interaction history, models each stream individually and efficiently with reduced semantic interference.
This design mitigates context pollution in long and heterogeneous user histories and yields consistent gains over strong sequential baselines in large-scale CTR prediction and retrieval tasks across multiple surfaces, while remaining practical for production-scale deployment.

\section*{Acknowledgments}
This work would not be possible without work from the following
contributors:
Haicheng Wang,
Haoyue Tang,
Yan Li,
Zefeng Zhang,
Keke Zhai,
Tony Chen,
Sharon Zhang,
Yujia Hao,
Trevor Waite,
Nan Zhang,
Bugra Akyildiz,
Renfei Chen,
Yue Dong,
Shiling Ding
Scott Dobbins,
Cao Gao,
Daisy Shi He,
Xinyao Hu,
Yanzun Huang,
Li Chen,
Justin Khim,
Li Lu,
Sophia (Xueyao) Liang,
Chloe Liu,
Xingyu Liu,
Min Ni,
Yiyi Pan,
Dawei Sun,
Hao Wang,
Ning Wang,
Xiaodong Wang,
Zhe (Joe) Wang,
Bi Xue,
Siqi Yan,
Jiyuan Zhang,
Lu Zhang,
Yuting Zhang,
Wei Zhao,
Chuanhao Zhuge,
Mingda Li,
Jeff Wang,
Honghao Wei,
Jeff Peng,
Haomin Yu,
Ke Pan.

\bibliographystyle{assets/plainnat}
\balance{}
\bibliography{sample-base}

\end{document}